\documentclass[twocolumn,showpacs,superscriptaddress,amsmath,amssymb]{revtex4}

\usepackage{dcolumn}
\usepackage{graphicx}

\begin{document}

\title{Effect of Zn and Ni impurities on the quasiparticle renormalization in Bi-2212}

\author{V.~B.~Zabolotnyy}
\affiliation{Institute for Solid State Research, IFW-Dresden,
P.O.Box 270116, D-01171 Dresden, Germany}

\author{S.~V.~Borisenko}
\affiliation{Institute for Solid State Research, IFW-Dresden,
P.O.Box 270116, D-01171 Dresden, Germany}

\author{A.~A.~Kordyuk}
\affiliation{Institute for Solid State Research, IFW-Dresden,
P.O.Box 270116, D-01171 Dresden, Germany} \affiliation{Institute
of Metal Physics of National Academy of Sciences of Ukraine, 03142
Kyiv, Ukraine}

\author{J.~Fink}
\affiliation{Institute for Solid State Research, IFW-Dresden,
P.O.Box 270116, D-01171 Dresden, Germany}

\author{J.~Geck}
\author{A.~Koitzsch}
\author{M.~Knupfer}
\author{B.~B\"{u}chner}
\affiliation{Institute for Solid State Research, IFW-Dresden,
P.O.Box 270116, D-01171 Dresden, Germany}

\author{H.~Berger}
\affiliation{Institute of Physics of Complex Matter, EPFL, CH-1015
Lausanne, Switzerland}

\author{A.~Erb}
\address{Walther-Mei\ss ner-Institut, Bayerische Akademie der Wissenschaften,
 Walther-Mei\ss ner Strasse 8, 85748 Garching, Germany}

\author{C.~T.~Lin}
\address{Max-Planck Institut f\"ur Festk\"orperforschung, D-70569 Stuttgart, Germany}

\author{B.~Keimer}
\address{Max-Planck Institut f\"ur Festk\"orperforschung, D-70569 Stuttgart, Germany}

\author{R.~Follath}
\affiliation{BESSY GmbH, Albert-Einstein-Strasse 15, 12489 Berlin,
Germany}


\begin{abstract}
The Cu substitution by Zn and Ni impurities and its influence on
the mass renormalization effects in angle resolved photoelectron
spectra (ARPES) of Bi$_2$Sr$_2$CaCu$_2$O$_{8-\delta}$ is
addressed. We show that the nonmagnetic Zn atoms have much
stronger effect both in nodal and antinodal parts of the Brillouin
zone than magnetic Ni. The observed changes are consistent with
the behaviour of the spin resonance mode as seen by inelastic
neutron scattering in YBCO. This strongly suggests that the
``peak-dip-hump'' and the ``kink" in ARPES on the one side and
neutron resonance on the other are closely related features.
\end{abstract}


\preprint{\textit{xxx}}

\maketitle The unique sensitivity of the angle-resolved
photoemission (ARPES) to the many-body effects in solids brings
this technique to the forefront of the modern research in the
field of high-temperature superconductivity (HTSC). The anomalies
in the single-particle spectral function of a superconductor,
detected and well characterized along the high symmetry
directions, are commonly believed to be crucial for understanding
HTSC. Along the nodal direction the renormalization effects are
represented by the unusual dispersion, the so called ``kink"
\cite{VallaScience,BogdanovPRB200,KaminskiPRL2001}. In the
vicinity of ($\pi$, 0) point of the Brillouin zone (BZ), where the
order parameter reaches its maximum, the renormalization is
noticeably stronger and makes itself evident even in the lineshape
of the single energy distribution curve (EDC)\cite{NormanPRL1997,
GromkoPRB2003,KimPRL2003}. It is widely accepted that coupling to
a collective mode\;\cite{Engelsberg1963, SandvikPRB2004} naturally
explains these anomalies. The origin of this mode remains a
current controversy between two most frequently proposed
candidates that are phonons and magnetic excitations.

Existing photoemission data suggest substantial dependence of the
coupling between the electrons and the mode on
momentum\;\cite{BogdanovPRB200}, hole doping \;\cite{KimPRL2003,
JohnsonPRB2001}, and temperature\;\cite{GromkoPRB2003,
KordyukPRL2004}, which well fits the magnetic scenario. However,
other similar ARPES experiments are interpreted in terms of
phonons\;\cite{CukPRL2004, DevereauxPRL2004}. In order to resolve
this controversy one needs a way to vary separately either the
phononic or magnetic spectrum. Variation in the phononic spectrum
can be achieved by the use of different isotopes. Such an
experiment for HTSC has recently been
reported\;\cite{LanzaraNature2004} but its ``unusual" character
(all major effects have been observed at high energies,
100-300\;meV) leaves the problem of low-energy anomalies, which
are relevant for the superconductivity, unsolved. In case of
magnetic excitations, there is a possibility to change the
magnetic spectrum via doping different types of impurities into
the Cu-O plane. It is known from inelastic neutron scattering
(INS)\;\cite{SidisPRL2000, SidisCodMat} that substitution of Zn
and Ni leads to substantial changes in the magnetic spectrum.

In this paper we show that substitution of Zn and Ni essentially
influences the renormalization anomalies in ARPES spectra both
along the nodal and antinodal regions of the Brillouin zone.
Moreover, the changes in ARPES spectra are in good correspondence
to those observed in magnetic spectra by INS.  This provides
evidence that it is the magnetic excitations that are responsible
for the unusual renormalization features in the quasiparticle
excitation spectrum of cuprates.

The ARPES experiments were carried out at the U125/1-PGM beam line
with an angle-multiplexing photoemission spectrometer (SCIENTA
SES100) at BESSY. Overall energy resolution was set to 20 meV, and
angular resolution was 0.2$^\circ$, which yields a wave vector
resolution 0.013\;{\r{A}}$^{-1}$ and 0.009\;{\r{A}}$^{-1}$ for
50\;eV and 27\;eV excitation energies, respectively. All data were
acquired at $T$\;=\;30\;K in the superconducting state for three
high quality samples: nearly optimally doped pure BSCCO ($T_C$ =
92 K), Zn substituted BSCCO with nominal Zn concentration close to
1\% ($T_C$\;= 86\;K), and 2\%\;Ni substituted BSCCO ($T_C$ = 87
K).

We focused on two regions of the BZ, the nodal and the antinodal,
which are schematically shown in a Fig.1 a by the two
double-headed arrows. Here the solid red line stands for the bare
dispersion of the antibonding band and the dashed blue one for the
bonding band\;\cite{KordyukPRB2003}.

\begin{figure}
\begin{center}
\includegraphics[width=6.8cm]{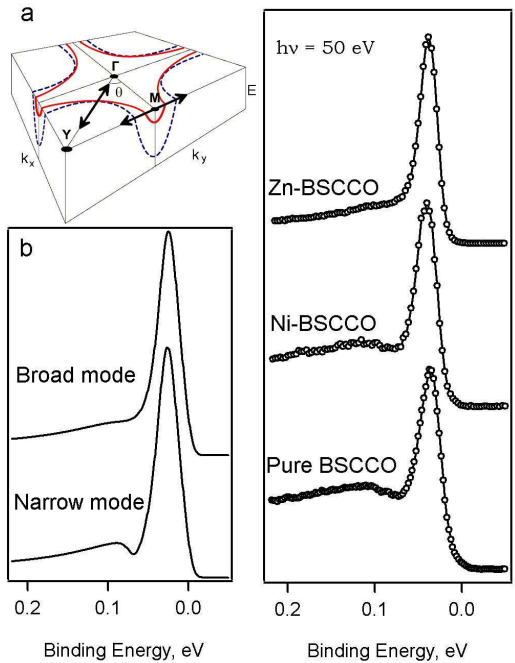}\\
\caption{\label{Images}(color). (a) Cut of a $k\omega$-space at
the Fermi level. The top of the cube contains two sheets of the
Fermi surface, red solid line corresponds to the dispersion of the
antibonding band, blue dashed one --- to the bonding band. $k_x$
and $k_y$ axes are directed along the horizontal edges of the
cube. (b) Model line shape of the $(\pi,0)$ EDC, when coupled to a
narrow (2\;meV) and broad (20\;meV) collective mode. (c)
Experimental EDC taken at $(\pi,0)$ point.}
\end{center}
\end{figure}

First we consider the antinodal region. In general case the EDC,
the curve that represents distribution of photo-intensity as a
function of energy at fixed momentum, taken at  the ($\pi$, 0)
point (see panel a Fig 1)  would consist of  two peaks due to the
bilayer split band\;\cite{KordyukPRL2002}. It has been shown that
the contribution from the bonding band can be completely removed
by a proper choice of excitation energy \cite{KordyukPRL2002}(in
this case $h\nu$ = 50\;eV). Nevertheless, at least for the
underdoped samples\;\cite{BorisenkoPRL2003}, the line shape is not
reduced to a single peak and has a dip at a higher binding energy.
The appearance of the dip in a single band EDC is well explained
in terms of coupling to a collective mode\;\cite{Engelsberg1963,
SandvikPRB2004}. Such an interaction with a sharp mode with energy
$\Omega_{mode}$ leads to the increase of the scattering rate at
binding energies higher than $\Omega_{mode}+\Delta$, where
$\Delta$ is a superconducting gap, i.e. to a step like feature in
the imaginary part of the self-energy and a logarithmic divergence
in the real part of the self-energy giving rise to a dip in
photoemission line shape. When the mode has a finite width the dip
smears out as demonstrated in Fig. 1 b. Therefore, the dip can be
a good indicative of the mode width and energy. Broadening of the
mode described above well correspond to our experimental data
presented in panel (c) of Fig. 1. Comparing these EDC's, measured
at ($\pi$, 0) point with 50\;eV photon energy, we see a striking
difference in the lineshapes caused by the impurity substitution:
while for the pure sample the spectrum has a pronounced dip, for
the Zn substituted sample the dip almost completely vanishes. For
the Ni substituted sample the dip ``strength" is intermediate. The
dip vanishing with Zn doping is a robust effect that we observed
for numerous samples and was one of the reasons to conduct this
systematic study.

Another change in the spectra that one can notice is an increase
of the intensity of the low-energy peak relative to the spectral
weight at higher binding energies upon Ni substitution and even
stronger effect on Zn substitution. In a simple model where we
explain the spectral weight at the ($\pi$,0) point by a strong
coupling to a mode (e.g. the neutron resonance mode) such an
intensity variation could be accounted for by a stronger coupling
to the mode upon impurity substitution.  A higher coupling to the
mode would be related to a larger imaginary part of the
self-energy and this would cause a stronger broadening and
therefore a reduction of the intensity in the ``hump" region. On
the other hand such a broadening would also result in a flattening
of the hump that we do not observe here. In addition, we do not
see in the dispersion of the bonding band (not shown) any clear
evidence for an increase of the mass enhancement with Zn doping.
Thus, in order to account for all changes in the line shape one
needs a model that also takes into consideration possible
difference in elastic scattering contributions (e.g. due to
different surface qualities) and different extrinsic backgrounds.

\begin{figure}[!b]
\begin{center}

\includegraphics[width=7.9cm]{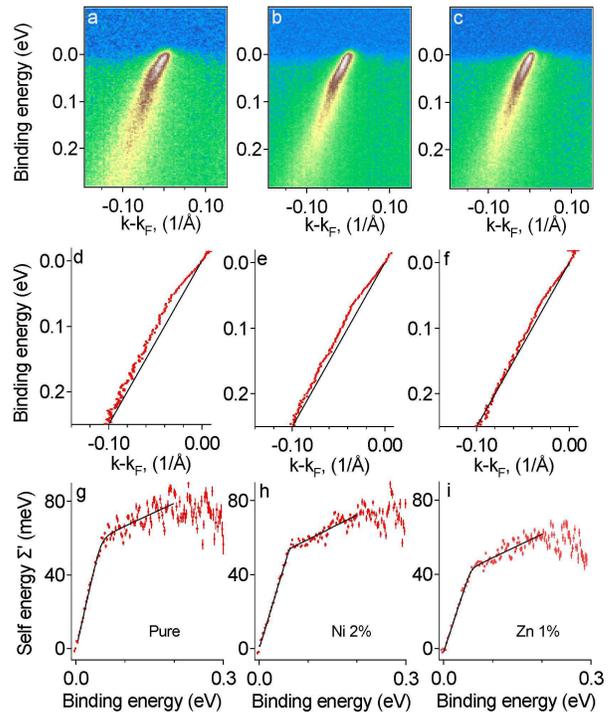}\\
\caption{(color). (a)-(c) Photo intensity distribution in the
nodal cuts. (a) Pure BSCCO, (b) Ni substituted and (c) Zn
substituted BSCCO. Middle row shows corresponding experimental
dispersion. In the  last row the real part of the self-energy
obtained by subtractions of the bare band is displayed. The black
solid lines are guides to eye, while the red symbols represent the
data points.}
\end{center}
\end{figure}

Now we turn to the nodal direction The nodal cuts (Fig. 2, a-c)
were measured strictly along the BZ diagonal. The aforementioned
position in \emph{k}-space was guaranteed by the measured maps. In
addition it was double checked using the steepest experimental
dispersion as a criterion for the true nodal cut. Since the nodal
region is also complicated by bilayer
splitting\;\cite{KordyukPRB2004}, the matter of excitation energy
remains important here too. The nodal spectra were measured with
$h\nu$ = 27 eV, when again only one of the bilayer split bands is
visible\;\cite{KordyukPRB2004}. In the second row of Fig. 2 we
show the experimental dispersion obtained from the maxima position
of the Lorentzians that provide the best fit to the momentum
dispersion curves (MDC). The straightforward way to estimate a
degree of the renormalization effects is to draw one and the same
``guide" line from the Fermi level to some fixed point on the
experimental dispersion curve and treat the area between the
dispersion curve and the guide line as an effective measure of
renormalization. It is easy to see that renormalization in case of
the Zn substituted sample is decreased in comparison to the pure
sample. Subtracting a bare particle band from the experimental
dispersion the real part of the self-energy can be obtained. Here,
as follows from a self-consistent Kramers-Kr\"onig procedure
\cite{KordyukPRB2005}, for the bare band dispersion we used  a
parabolic band with its bottom lying 0.8 eV below the Fermi
level(FL). Defining a coupling constant as derivative of the real
part of the self-energy at the FL we find that
$\lambda_{pure}\approx1.06$, $\lambda_{Ni}\approx0.88$, and for
the Zn doped sample it has the lowest value
$\lambda_{Zn}\approx0.76$. We note that, although the obtained
absolute values for $\lambda$ depend on the parameters of the bare
band dispersion, their ratio remains almost insensitive to the
position of the bottom of the bare band, provided that the bare
band is the same for all three samples. Therefore more accurate
would be to say that the coupling constant for the nodal direction
is reduced by approximately 15\% for the 2\% Ni doped sample and
by 30\% for 1\% Zn doped sample as compared to pure one.

\begin{figure}[!b]
\begin{center}
\includegraphics[width=7.9cm]{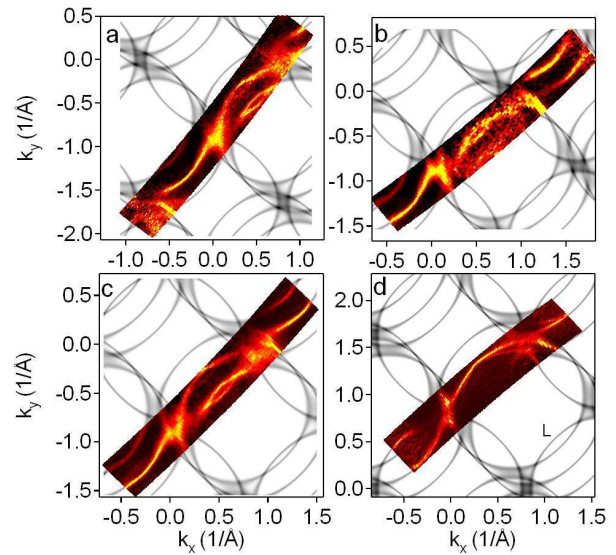}
\caption{(color). Cuts of $k\omega$-space made 20\;meV below the
Fermi level and tight binding fits. (a) Ni substituted sample. (b)
Zn substituted sample. (c) Pure BSCCO close to optimal doping. (d)
Overdoped BSCCO, $\delta\approx0.20$; letter L denotes ``lens"
formed by the shadow band and the 1st order diffraction replica.}
\end{center}
\end{figure}

A link between the modifications in our spectra and changes in the
spectrum of magnetic excitations with impurity substitution was
based on the implication that we have equal charge doping of the
samples. It is known that the renormalization features are
weakening with overdoping both in nodal and antinodal directions
\cite{JohnsonPRB2001,KimPRL2003}. Therefore different hole doping
might be a possible explanation. In order to evaluate the hole
doping level we measured the maps in \emph{k} space for all
samples, which, in addition, allowed us to know precisely our
position in the reciprocal space when measuring nodal and
antinodal spectra. In a false colour scale (Fig.3) we show
experimental data, which are the cuts of $k\omega$-space made
20\;meV below the Fermi level for the three samples under
consideration (a - c). For comparison, in the  panel d we give the
identical cut for overdoped sample with the hole doping $\delta$
within 0.19 -- 0.20. Grey scale images below the experimental data
represent tight-binding fits to the antibonding sheet of the Fermi
surface (FS)\;\cite{Andersen1995}. We focus attention on the
``lens" (Fig. 3 d) formed by the shadow and the 1st order
diffraction replica. Even from its size, which is determined by
the size of the FS, it is clear that the first three samples have
very close hole doping, and that the doping of the fourth sample
is notably higher. To be quantitative, the area of antibonding
sheets of the FS cuts can be used as a good relative doping
measure. Evaluation gives the following result:
$(S_a-1/2):(S_b-1/2):(S_c-1/2):(S_d-1/2)\approx$
1.00\;:\;1.01\;:\;0.98\;:\;1.14, where $S_i$ denotes the ratio of
the area confined by one ``barrel" to the area of the whole BZ.
The values for $S_i$  were obtained from the tight-binding fit of
experimental map for each panel (a-d). However, a charge doping
induced suppression of the renormalization features would require
a deviation of the hole concentration of about 30 \%
\;\cite{KimPRL2003}. So we see that the difference in charge
doping for our sample is negligible to account for substantial
change in the line shape (Fig. 1 c) at the antinode, as well as
for the different renormalization constants (Fig. 2) for the nodal
spectra.

We find that our experimental observations of the dip smearing in
the antinodal EDC are in good agreement with the results of
inelastic neutron scattering (INS) experiments. The comparison
with INS is most pertinent here, since it gives a detailed insight
into the changes in the spectrum of magnetic excitations produced
by the impurities and allows to make a comparison with those that
we see in ARPES spectra. The direct observation of the magnetic
resonance mode by INS in YBCO demonstrates its strong sensitivity
to impurities incorporated into Cu-O planes. When for pure sample
the mode is sharply peaked at $\sim$ 40\;meV with a FWHM being
resolution limited \cite{FongPRL1995}, for 1\% Zn substituted YBCO
the mode attains finite width of about 10\;meV\cite{SidisPRL2000}.
Gradual broadening of the resonance and decrease  of its peak
intensity is traced in details in Ref.~\onlinecite{SidisCodMat}.
From Ref.~\onlinecite{SidisPRL2000} it follows that introduction
3\% of Ni result in approximately the same intensity of the
resonance at its peak as in the case of 1\% Zn doping, supporting
the idea that nonmagnetic Zn has stronger influence on spin
dynamics than Ni, which is the same that we see in ARPES spectra.

Raman scattering experiments on optimally doped YBCO
\cite{TaconCodMat} show that the A$_{1g}$ electronic collective
mode, that follows the acoustic part of magnetic resonance,
gradually broadens with increase of Zn concentration. For 2\% Zn
substituted YBCO the A$_{1g}$ mode almost disappears, while this
is not the case for 3\% Ni substituted YBCO where the mode loses
only about half of its intensity \cite{GallaisPRL2002}. It is also
worth mentioning that Raman scattering experiments on Zn and Ni
substituted YBCO crystals show no apparent changes in the
structure of phonon anomalies \cite{Watanabe1998,Kall1996} at room
temperature suggesting that the lattice dynamics remains
unchanged. Similar results, except for the peak at 340\ cm$^{-1}$,
are obtained in Ref. \onlinecite{Limonov2001} for superconducting
state. This peak shows small decrease of its intensity and
insignificant, as compared to magnetic resonance mode, broadening
from 1.6 to 1.9\;meV. It is clear that such a broadening cannot
account for the vanishing of the dip in the $(\pi, 0)$ spectra.
Therefore we conclude that the smearing of the dip in the APES
antinodal spectra with impurities doping is a straightforward
consequence of the magnetic resonance mode broadening, and
possibly decrease of its integral intensity.

For the nodal direction the situation is more complicated.
According to \cite{EschrigPRB2003} the renormalization effects in
nodal spectra are mainly caused by scattering of the electrons
between nodes and Van Hove singularity in the density of states at
$(\pi,0)$ point by the resonance mode centered at
antiferromagnetic vector $Q(\pi,\pi)$. This serves as a good
explanation for the  fact that the renormalization effects are
much weaker in the node than those in the vicinity of $(\pi,0)$
point. But it also means, that the renormalization has to increase
with the resonance mode broadening in momentum space when
impurities are doped into Cu-O plaquett\'{e} \cite{SidisPRL2000}.
On the contrary, our observations show that the dimensionless
coupling constant $\lambda$, decreases suggesting that the
scattering by the antiferromagnetic vector $Q(\pi,\pi)$ does not
have the dominant contribution in the nodal spectra
renormalization effects \cite{ChubukovPRB2004}. The new resonance
feature in magnetic excitation spectrum \cite{PailhesPRL2004,
EeminPRL2004} peaked at energy close to 54\;meV and incommensurate
vector $Q^*_{IC}(0.8\pi,0.8\pi)$ provides scattering between the
nodes, and therefore,  should have a strong influence on the nodal
spectra renormalization effects. However its evolution with
impurity doping needs to be clarified, in order to check its
pertinence for the problem in hand.

In conclusion, we have shown that the substitution of Cu atoms in
Cu-O plane changes renormalization features in ARPES spectra both
in nodal and antinodal parts of the Brillouin zone. The smearing
of the dip in the antinodal EDC can be well explained by coupling
of electrons to magnetic resonance mode. The effect of Zn and Ni
substitution on the antinodal ARPES spectra is in good agreement
with the influence of these impurities on magnetic resonance mode
seen by INS experiments. These in addition to the previous ARPES
studies of temperature and doping dependence
\cite{BogdanovPRB200,JohnsonPRB2001,KimPRL2003,KordyukPRB2004} of
PDH structure, mass renormalization near antinodal region and a
kink in a nodal part of Brillouin zone, provides another strong
evidence that the mode, coupling to which, causes observed
renormalization effects has rather magnetic than phononic origin.

The project is part of the Forschergruppe FOR538 and is supported
by the DFG under Grants No. KN393/4 and No. 436UKR17/10/04. We
acknowledge stimulating discussions with  I. Eremin, and technical
support by R.\;H\"{u}bel, S. Leger.

\end{document}